%% file: main.tex
\documentclass[ALICE,manyauthors]{cernphprep}
\usepackage[comma,square,numbers,sort&compress]{natbib}
\usepackage{hyperref}
\usepackage{lineno}
\usepackage{xspace} 
\usepackage[all]{nowidow} 
\usepackage{isomath} 
\usepackage{siunitx} 

\usepackage{bm} 
\usepackage{hyphenat} 
\usepackage{xcolor}
\usepackage{upgreek}
\usepackage{amsmath}
\usepackage[T1]{fontenc}
\usepackage{orcidlink}

\input{commands}

\begin{document}

\begin{titlepage}
\PHyear{2023}       
\PHnumber{265}      
\PHdate{17 November}  

\title{Emergence of long-range angular correlations in\linebreak low-multiplicity proton--proton collisions}
\ShortTitle{Long-range correlations in low-multiplicity proton--proton collisions}

\Collaboration{ALICE Collaboration\thanks{See Appendix~\ref{app:collab} for the list of collaboration members}}
\ShortAuthor{ALICE Collaboration} 

\begin{abstract}
\input{abstract}

\end{abstract}
\end{titlepage}

\setcounter{page}{2} 


\input{intro}

\label{Sec:intro}

\label{Sec:Setup}
\input{setup}

\label{Sec:Analysis_Procedure}
\input{analysis}


\label{Sec:Systematics}
\input{systematics}


\label{Sec:Results}
\input{results}

\input{conclusions}
\label{Sec:Conclusions}

\newenvironment{acknowledgement}{\relax}{\relax}
\begin{acknowledgement}
\section*{Acknowledgements}
\input{fa_2023-11-06_Opt_C.tex}
\end{acknowledgement}

\bibliographystyle{utphys}   
\bibliography{bibliography}

\newpage
\appendix

%
%

\section{The ALICE Collaboration}
\label{app:collab}
\input{2023-11-06-Alice_Authorlist_2023-11-06_Opt_C.tex}

\end{document}

%% file: commands.tex
\newcommand{\pp}           {pp\xspace}

\newcommand{\ee}           {\mbox{$\mathrm {e^{+}e^{-}}$}\xspace}

\newcommand{\pt}           {\ensuremath{p_{\mathrm{T}}}\xspace}

\newcommand{\Ncorr}        {\ensuremath{\langle N_\mathrm{ch}\rangle}\xspace}
\newcommand{\cpp}          {\ensuremath{c_{\mathrm{pp}}}\xspace}

\newcommand{\ceeA}          {\ensuremath{c_{\mathrm{ee}}^{1}}\xspace}
\newcommand{\ceeB}          {\ensuremath{c_{\mathrm{ee}}^{2}}\xspace}

\newcommand{\dEta}           {\ensuremath{\Updelta\eta}}
\newcommand{\dPhi}           {\ensuremath{\Updelta\varphi}}

\newcommand{\pttrig}         {\ensuremath{p_{\rm T,trig}}\xspace}
\newcommand{\ptassoc}        {\ensuremath{p_{\rm T,assoc}}\xspace}

\newcommand{\GeVc}         {\ensuremath{\mathrm{GeV}/c}\xspace}

\newcommand{\seven}     {\ensuremath{\sqrt{s}=7\,\mathrm{TeV}}\xspace}
\newcommand{\thirteen}     {\ensuremath{\sqrt{s}=13\,\mathrm{TeV}}\xspace}
\newcommand{\nineone}      {\ensuremath{\sqrt{s}=91\,\mathrm{GeV}}\xspace}
\newcommand{\alephhighenergy} {\ensuremath{\sqrt{s}=183\text{--}209\,\mathrm{GeV}}\xspace}

%% file: abstract.tex
This Letter presents the measurement of near-side associated per-trigger yields, denoted ridge yields, from the analysis of angular correlations of charged hadrons in proton--proton collisions at \thirteen. Long-range ridge yields are extracted for pairs of charged particles with a pseudorapidity difference of $1.4 < |\Updelta\eta| < 1.8$ and a transverse momentum of $1 < \pt < 2\,\GeVc$, as a function of the charged-particle multiplicity measured at midrapidity. This study extends the measurements of the ridge yield to the low multiplicity region, where in hadronic collisions it is typically conjectured that a strongly-interacting medium is unlikely to be formed. The precision of the new low multiplicity results allows for the first direct quantitative comparison with the results obtained in \ee collisions at \nineone and \alephhighenergy, where initial-state effects such as pre-equilibrium dynamics and collision geometry are not expected to play a role. In the multiplicity range $8\lesssim\langle N_\mathrm{ch}\rangle\lesssim 24$ where the \ee results have good precision, the measured ridge yields in pp collisions are substantially larger than the limits set in \ee annihilations.
Consequently, the findings presented in this Letter suggest that the processes involved in \ee annihilations do not contribute significantly to the emergence of long-range correlations in pp collisions.

%% file: intro.tex
Long-range angular correlations have been observed in the collisions of ultra-relativistic heavy ions both at RHIC~\cite{Adams:2005dq,Adcox:2004mh,Arsene:2004fa,Back:2004je} and LHC~\cite{Abelev:2012di, Abelev:2014pua, ATLAS:2011ah}. In such collisions, the presence of correlations between pairs of particles with large pseudorapidity differences is interpreted as a signature for the existence of a strongly-coupled medium, known as the quark--gluon plasma (QGP), which converts the initial pressure gradients created in non-central nucleus--nucleus collisions into a collective momentum anisotropy of the final-state hadrons. In particular, the collectivity signal manifests itself as a near-side ridge around the jet fragmentation peak in the two-particle correlation function, indicating the presence of a medium with considerable anisotropic flow.

Long-range correlations have also been observed in high-multiplicity proton--proton (pp)~\cite{CMS:2010ifv,ATLAS:2015hzw,Khachatryan:2015lva,Khachatryan:2016txc,Acharya:2019vdf,ATLAS:2017rtr}, proton--nucleus (pA)~\cite{ALICE:2012eyl,ATLAS:2014qaj,ATLAS:2016yzd,Khachatryan:2016ibd}, and light nucleus--nucleus collisions~\cite{PHENIX:2018lia,Aidala:2017ajz,STAR:2015kak}. These results have challenged the interpretation of the so-called collective phenomena in hadronic collisions~\cite{Voloshin:2008dg}, and raised the question whether the same underlying dynamics can be responsible for the emergence of long-range correlations in small and large systems~\cite{Nagle:2018nvi}. Notably, the formation of a medium and its subsequent evolution, which is understood to take place in heavy-ion collisions might not be justifiable in small collision systems, where the requirement of thermal equilibrium may not be achieved under the conditions of small system size.
Despite a vast experimental and theoretical effort, an unambiguous description of these experimental data is not yet achieved~\cite{Strickland:2018exs,Loizides:2016tew,Nagle:2018nvi}, although there has been recent progress~\cite{Schenke:2020mbo,Zhao:2022ugy,STAR:2022pfn}. Flow-like signatures could indeed originate from the very early stages of the collisions~\cite{Dusling:2012cg,Bzdak:2013zma} or develop during the late stages of the collisions as a consequence of the interaction with a strongly-coupled medium~\cite{Greif:2017bnr,Mantysaari:2017cni}. This suggests that ridge measurements combined with measurements of jet shape modification can eventually be used to separate initial state and flow-driven contributions~\cite{Schenke:2022mjv}.

Recently, new experimental insights have been obtained by the study of long-range correlations in \ee collisions at \nineone~and \alephhighenergy~using ALEPH~\cite{DECAMP1990121} archived data~\cite{Badea:2019vey,Chen:2023njr}.
Collisions between point-like electrons and positrons remain unaffected by the presence of beam remnants or gluonic initial-state radiation, and are not sensitive to the modeling of parton distribution functions~\cite{PhysRevC.97.024909,Castorina:2020iia}.
Near-side ridge is neither observed in the lab reference frame nor in the thrust-axis reference frame. The results obtained in \ee collisions were also compared with the associated yield measurement in pp collisions with CMS~\cite{CMS:2015fgy,CMS:2010ifv},
but due to the large uncertainties of the existing pp measurement, a statistically significant comparison between the ridge yields measured in pp and \ee collisions was not feasible. 
A systematic study of such signatures across collision systems and sizes represents a unique opportunity to characterize the emergence of collective phenomena. In particular, measurements performed in pp collisions with very low multiplicity can provide crucial inputs to address the relevance of initial-state effects~\cite{Greif:2017bnr} in the presumed absence of a flow-inducing medium and final-state correlations~\cite{OrtizVelasquez:2013ofg,Bierlich:2022pfr}, and in turn constrain the magnitude of these initial-state effects traditionally afflicted by a large uncertainty~\cite{Song:2010mg}. At the same time, comparing to the \ee collision system helps to identify physical processes in the pp system that do contribute to collectivity, as no such signals were detected in the \ee system~\cite{Badea:2019vey,Chen:2023njr} up to certain multiplicities.

In this Letter, the near-side long-range yields are measured in pp collisions at \thirteen with good precision down to very low multiplicities. The results are reported for pairs of charged particles with pseudorapidity $1.4<|\dEta|<1.8$ and transverse momentum $1<\pt<2\,\GeVc$, as a function of the charged-particle multiplicity measured at midrapidity. The experimental precision of this study allows for the first quantitative comparison with the results obtained in \ee collisions at \nineone and \alephhighenergy with ALEPH archived data, where initial-state effects in hadronic collisions are not expected to play a role. The measurement is also compared to predictions of PYTHIA 8.3~\cite{Bierlich:2022pfr} and EPOS LHC simulations~\cite{Pierog:2013ria}.

%% file: setup.tex
The data were collected in 2017 and 2018 using the ALICE apparatus~\cite{ALICE:2022wpn} at the LHC. Information about the detector configuration and performance can be found in Refs.~\cite{Aamodt:2008zz,ALICE:2014sbx}. The main systems used for this study were the central\hyp{}barrel detectors, located 
within a~solenoidal magnet and used for charged-particle tracking.
These include the Inner Tracking System and the Time Projection Chamber.
The V0 detector, consisting of two scintillator arrays located at pseudorapidities of $2.8 < \eta < 5.1$ and $-3.7 < \eta < -1.7$, was used for the trigger and event selections. Minimum\hyp{}bias (MB) events were selected online by requiring a~signal from at least one charged particle in both V0 counters.
The analyzed data sample consists of about $1.3 \times 10^9$ MB \pp{} collisions at a center-of-mass energy of \thirteen within primary vertex range $|z_\mathrm{vtx}|<8.0\,\mathrm{cm}$ along the beam axis, corresponding to an integrated luminosity of about $22\,\mathrm{nb}^{-1}$.

%% file: analysis.tex
The observables presented in this analysis are extracted using  two-particle angular correlations for pairs of charged particles. The two-particle per-trigger yield is measured as a function of relative azimuthal angle $\Updelta\varphi$ and pseudorapidity $\Updelta\eta$ of two particles --- traditionally called trigger and associated --- and is defined as
\begin{equation}
\label{eq:twocorr}
\frac{1}{N_{\mathrm{trig}}}\frac{\mathrm{d}^{2}N_{\mathrm{pair}}}{\mathrm{d}\Updelta\eta\mathrm{d}\Updelta\varphi} = B(0,0)\frac{S(\Updelta\eta,\Updelta\varphi)}{B(\Updelta\eta,\Updelta\varphi)}.
\end{equation}
Equation~\eqref{eq:twocorr} is evaluated within a range of transverse momentum of the trigger ($\pttrig$) and associated ($\ptassoc$) particles within $|\eta|<1.0$ where $\pttrig > \ptassoc$. The total number of trigger particles is denoted with $N_\mathrm{trig}$, and the number of trigger and associated particle pairs with $N_\mathrm{pair}$. This two-particle yield $S(\Updelta\eta,\Updelta\varphi)$ is corrected for pair acceptance and reconstruction effects by constructing a mixed-event distribution $B(\Updelta\eta,\Updelta\varphi)$ from pairs where the trigger and associated particles are taken from different events. This mixed-event distribution is normalized by $B(0,0)$ computed by pairs of particles of identical charge traveling in the same direction for which acceptance and reconstruction effects are identical by construction. The event mixing is performed such that events with similar multiplicity and primary vertex $z_\mathrm{vtx}$ (bins of 2 cm) are combined. The final per-trigger yield is obtained by averaging over these individual bins. In addition, all tracks are corrected for the single-particle tracking efficiency as a function of $p_\mathrm{T}$ and $\eta$. The efficiency corrections and acceptance factors are obtained by simulating events with PYTHIA 8.3 with the Monash tune~\cite{Skands:2014pea}, and the detector response simulated using the GEANT3 transport package~\cite{Brun:1994aa}.

The per-trigger yield distribution as a function of $\Updelta\varphi$ is obtained by integrating the two-dimensional two-particle per-trigger yield in the long-range intervals $1.4<|\Updelta\eta|<1.8$, in order to exclude the region dominated by the jet fragmentation peak
\begin{equation}
\label{eq:pertrigg}
Y(\Updelta\varphi) = \frac{1}{N_{\mathrm{trig}}}\frac{\mathrm{d}N_{\mathrm{pair}}}{\mathrm{d}\Updelta\varphi} = \int_{1.4<|\Updelta\eta|<1.8}\left(\frac{1}{{N}_{\mathrm{trig}}}\frac{\mathrm{d}^{2}N_{\mathrm{pair}}}{\mathrm{d}\Updelta\eta\mathrm{d}\Updelta\varphi}\right) \dfrac{1}{\delta_{\Updelta\eta}} \mathrm{d}\Updelta\eta,
\end{equation}
where $\delta_{\Updelta\eta}=0.8$ is the normalization constant for the chosen $\Updelta\eta$ range.
The ridge yield $Y^\mathrm{ridge}$ is extracted by integrating the near-side area of the associated per-trigger yield using
\begin{equation}
\label{eq:ridgeyield}
	Y^\mathrm{ridge}=\int_{|\Updelta\varphi|<|\Updelta\varphi_\mathrm{min}|}Y(\Updelta\varphi)\mathrm{d}\Updelta\varphi-2|\Updelta\varphi_{\min}|C_\mathrm{ZYAM}.
\end{equation}

A Zero-Yield-At-Minimum (ZYAM) procedure is applied to subtract the baseline of the per-trigger yield.
We assume that $Y(\Updelta\varphi)$ has an uncorrelated flat contribution, $C_\mathrm{ZYAM}=Y(\Updelta\varphi_{\min})$, where $\Updelta\varphi_{\min}$ is the location of the minimum of $Y(\Updelta\varphi)$. To reduce the impact of statistical fluctuations, $Y(\Updelta\varphi)$ is fitted by a symmetric Fourier series up to third harmonic $F(\Updelta\varphi) = \sum_{n=0}^3 2 a_n \cos (n\Updelta\varphi)$, which is found to be adequate in this analysis for a precise extraction of $\Updelta\varphi_{\min}$ and $C_\mathrm{ZYAM}$. This procedure is illustrated in the right panel of Fig.~\ref{fig:ratiosubtracted}.
Since the fit is only used to locate the bounds of the near-side ridge, the impact of the higher harmonics on the baseline and ZYAM procedure is negligible.
The ridge yield provides a measure for collective effects, and is generally compatible with the measurements of flow coefficients $v_n$ in small systems~\cite{ALICE:2023lyr}. The measurement of $Y^\mathrm{ridge}$ also facilitates the comparison with the readily available \ee result, and does not suffer from ambiguities related to low-multiplicity template subtraction applied in other measurements~\cite{ATLAS:2016yzd}.

The analysis is performed in different intervals of measured multiplicity. In order to determine the corrected charged-particle multiplicity $\Ncorr$ in each multiplicity interval, the number of charged tracks is counted within $|\eta| < 1$ and $\pt > 0.2\,\GeVc$.
This number is corrected for detector effects by correlating reconstructed and simulated multiplicities, and randomly sampling a new multiplicity value from the simulated distribution corresponding to the reconstructed value, representing the uncorrected measured multiplicity.
At the same time, the resampling technique reduces self-correlation between the multiplicity and the particles entering the per-trigger yield. The analysis is carried out for 14 multiplicity intervals, ranging from $N_\mathrm{ch}=0$ to 62, where the average MB multiplicity is about 11.3.

%% file: systematics.tex
The systematic uncertainties of the ridge yields are evaluated by varying the event and track selections as well as the integration ranges used in the extraction.
A bootstrapping procedure~\cite{Badea:2019vey}
is used to estimate both the statistical and systematic uncertainties. This uncertainty is obtained by making large number of variations of the default value $Y_\mathrm{def}(\Updelta\varphi)$ of the per-trigger yield distribution, by adding random statistical and systematic fluctuations and extracting the ridge yield applying the procedure as given in Eq.~\eqref{eq:ridgeyield}. Gaussian fluctuations are randomly added bin-by-bin based on the statistical uncertainty. Systematic fluctuations are included by assuming that the $Y_s(\Updelta\varphi)/Y_\mathrm{def}(\Updelta\varphi)$ variation for each source of systematic uncertainty $s$ has a common shift across $(\Updelta\varphi,\Updelta\eta)$ (not affecting the ridge yield) and a bin-by-bin component, taken to be Gaussian distributed and each variation corresponding to $1\sigma$. It should be noted that the statistical uncertainties and the five sources of systematic uncertainty described below are all varied each time. The final uncertainty on $Y^\mathrm{ridge}$ is calculated as the standard deviation of the yield distribution obtained from a large number of these random variations. For multiplicity intervals where the result is consistent with zero, a limit $Y_\mathrm{CL}^\mathrm{ridge}$ from this distribution is estimated at 95\% confidence level~(CL).

The selection on the position of the primary vertex along the beam axis ($z_\mathrm{vtx}$) is varied from $|z_\mathrm{vtx}|<8\,\mathrm{cm}$ to 10\,cm. The corresponding systematic uncertainty was found to be less than 5\% depending on multiplicity. In order to estimate the bias due to the possible presence of jet-like correlations in $1.4<|\dEta|<1.8$, the definition of the long-range region was changed to $1.5<|\dEta|<1.8$. 
This change also estimates the effect of residual non-flow in the region in which the ridge yield is extracted.
As the near-side ridge yield decreases towards low multiplicity, the relative contribution from jet-like (non-flow) contribution to the systematic uncertainty increases from 3\% at high multiplicity to 22\% at $\langle N_\mathrm{ch}\rangle=15$, and dominates for lower multiplicity.
Uncertainties related to track reconstruction were estimated by varying the required number of hits in the ITS layers allowing a more uniform detector acceptance but larger contributions from secondaries, resulting in 3\%--10\% variation without a clear multiplicity dependence. Residual two-particle acceptance effects due to the finite accuracy of the event-mixing pair-acceptance correction, generally affecting the structure at long-range $\dEta$, have been estimated to be around 4\% on average.
This contribution has been evaluated by adjusting the per-trigger yield with a $\Updelta\varphi$-independent factor such that the away-side region is constant over $\dEta$.
As a further check on the extraction of the associated yield, the $\Updelta\varphi$ integration region is shifted with a resulting difference of about~1\%. The total systematic uncertainty resulting from the bootstrapping procedure is around 5\% at $\langle N_\mathrm{ch}\rangle=60$ and increases gradually to 25\% at $\langle N_\mathrm{ch}\rangle=15$.

%% file: results.tex
\begin{figure}[tbh!]
    \centering
    \includegraphics[width=0.49\textwidth]{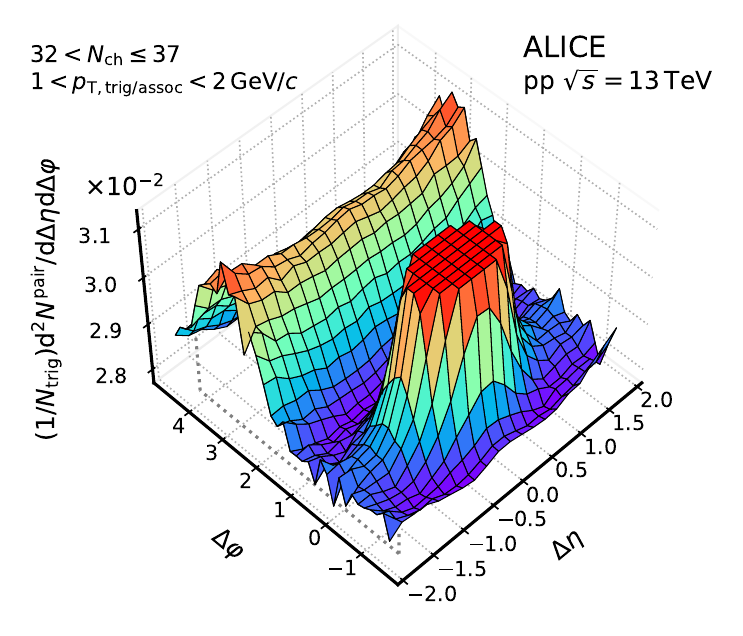}
    \includegraphics[width=0.49\textwidth]{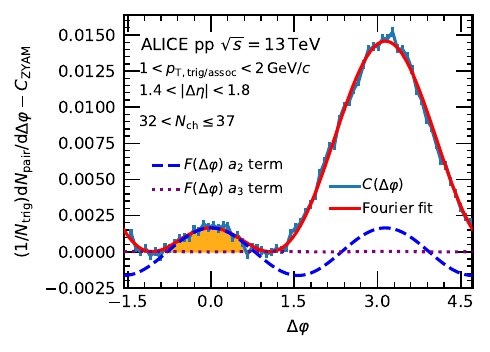}
    \caption{Two-particle per-trigger yield measured for charged track pairs with $1<\pttrig<2\,\GeVc$ and $1<\ptassoc<2\,\GeVc$ within the multiplicity range $32 < N_\mathrm{ch} \leq 37$. The jet fragmentation peak has been truncated to ensure a better visibility of the long-range structure. The right panel shows the zero-suppressed projection to $\dPhi$ overlaid with $F(\Updelta\varphi)$ (red line) and the area in which the ridge yield is extracted (shaded area). The blue and purple lines represent the second and third harmonic terms of $F(\Updelta\varphi)$.}
    \label{fig:ratiosubtracted}
\end{figure}

Figure~\ref{fig:ratiosubtracted} presents the two-particle per-trigger yield for trigger and associated particle momentum of \mbox{$1<\pt<2\,\GeVc$} in the multiplicity interval $32 < N_\mathrm{ch} \leq 37$. A prominent jet-fragmentation peak originating from correlations of particles from the fragmentation of the same parton is visible at $(\Updelta\eta,\Updelta\varphi)=(0,0)$. At $\Updelta\varphi=\pi$, a broad away-side structure results from correlations of tracks from back-to-back jet fragments that are spread over the entire $\Updelta\eta$ region (as the parton--parton scattering center-of-mass frame is not the lab frame). The momentum region has been chosen such that these peaks are sufficiently narrow in $\Updelta\varphi$, allowing one to extract the long-range ridge yield.
At $|\Updelta\eta|\gtrsim 1.4$ and $\Updelta\varphi\approx0$, the ``ridge'' structure, represented by an enhancement of the correlation, is visible which was observed in previous measurements~\cite{ALICE:2021nir} and that in heavy-ion collisions is interpreted as a sign of collective expansion of the QGP medium.
The overall shape and width of the jet fragmentation peak has been studied extensively to be confident that the non-flow contamination in the yield extraction is minimal. In addition, the $\Updelta\eta$ cuts have been varied as part of the systematic uncertainties. These investigations of the dependence of correlations on jet fragmentation and its implications on correlation studies are of interest in itself and will be published subsequently.

\begin{figure}[tbh!]
\centering
\hspace{-2.5em}\includegraphics[width=0.795\textwidth]{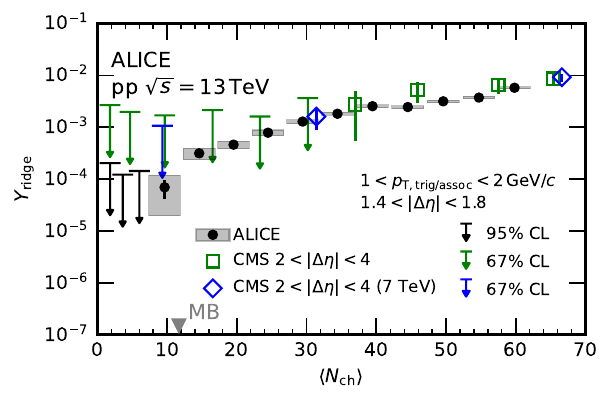}
	\caption{Ridge yield as a function of multiplicity. The black points correspond to the measurement presented in this Letter, while data from CMS~\cite{CMS:2010ifv,CMS:2015fgy} are drawn as green and blue markers. Vertical bars denote statistical uncertainties while systematic uncertainty is shown as shaded area. For both results, at low multiplicity where the lower uncertainty reaches zero, an upper limit is reported, which is drawn as a bar and down arrow. Such points are given at 95\% CL for the results from this Letter and at 67\% for the results from CMS. The ``MB'' arrow at $\langle N_\mathrm{ch}\rangle=11.3$ indicates the multiplicity averaged over the entire studied multiplicity range.}
\label{fig:ridgeyieldCMS}

\centering
\hspace{-2em}\includegraphics[width=0.81\textwidth]{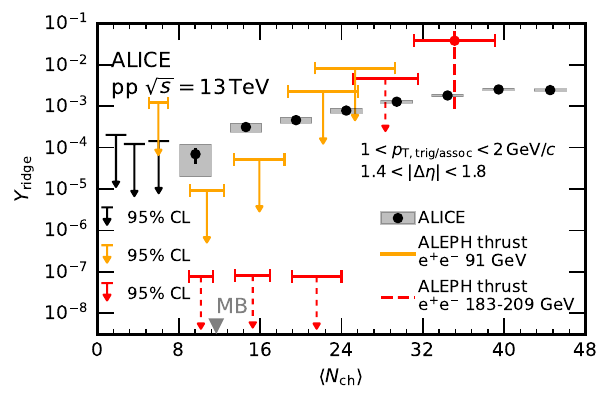}
\caption{Ridge yield as a function of multiplicity, compared to the upper limits on the ridge yield in \ee collisions. Vertical bars denote statistical uncertainties while systematic uncertainty is shown as the shaded areas. The orange limits represent the measurement in the thrust-axis reference frame with ALEPH~\cite{Badea:2019vey}. The horizontal bars in the ALEPH points represent the uncertainty related to the multiplicity conversion from the ALEPH to the ALICE acceptance (see text). All limits are given at 95\% CL.}
\label{fig:ridgeyieldALEPH}
\end{figure}

Figure~\ref{fig:ridgeyieldCMS} shows the extracted ridge yield $Y^\mathrm{ridge}$ as a function of the average charged-particle multiplicity. The measured $Y^\mathrm{ridge}$ shows a strong multiplicity dependence, with an increasing trend towards higher multiplicity collisions. A non-zero $Y^\mathrm{ridge}$ is measured with good precision for events with $\Ncorr > 9$, significantly extending the low-multiplicity reach of previous measurements~\cite{CMS:2015fgy}. A limit, represented in the figure by the black arrows, is computed at 95\% CL for the three lowest multiplicity intervals ($\Ncorr < 9$) where no significant ridge yield was observed.
The origin of the arrow corresponds to the threshold value of the ridge yield ($Y^\mathrm{ridge}_\mathrm{CL}$) for which 95\% of the bootstrap distribution values are smaller than $Y^\mathrm{ridge}_\mathrm{CL}$.
The results are compared with an analogous measurement performed by CMS (green markers) at the same center-of-mass energy. To allow for a direct comparison with the ALICE measurement, the $x$-axis of the CMS data was scaled by the ratio of the pseudorapidity acceptance of CMS and ALICE, which was estimated to be about 0.66 with negligible statistical uncertainty based on PYTHIA~8.3 simulations. The CMS result presents finite near-side yields for $\Ncorr \gtrsim 38$ and limits at 67\% CL for smaller multiplicities. The two results are in good agreement at high multiplicities, where an accurate estimation of the ridge yields is available for both experiments. The comparison also includes CMS measurements at \seven, with the same scaling procedure applied. The measurement at \seven has a smaller uncertainty at $\Ncorr\sim 32$ compared to the one at \thirteen and also agrees with the ALICE results.

\begin{figure}[tbh!]
\centering
\hspace{-2em}\includegraphics[width=0.81\textwidth]{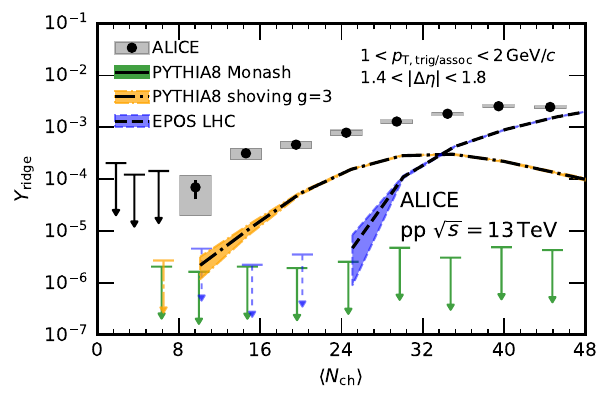}
\caption{Ridge yield as a function of multiplicity compared to the predictions of PYTHIA 8.3~\cite{Bierlich:2022pfr} with Monash tune~\cite{Skands:2014pea} (green) and string shoving~\cite{Bierlich:2017vhg} (orange) as well as EPOS LHC simulations~\cite{Pierog:2013ria} (blue). Because of a larger jet fragmentation peak width in the simulations than in data, the yield is extracted within $2<|\dEta|<4$ for the model calculations. A 95\% CL is indicated for model calculations when the lower limit of statistical uncertainty is below zero. Some points are slightly displaced along the $x$-axis for better visualization. The band indicates the statistical uncertainty from the event generation and the extraction procedure.}
\label{fig:ridgemodels}
\end{figure}

In Fig.~\ref{fig:ridgeyieldALEPH}, the result is compared to a recent measurement performed in \ee collisions at \nineone~\cite{Badea:2019vey} and at \alephhighenergy~\cite{Chen:2023njr} in the thrust-axis reference frame using ALEPH archived data.
Because of the absence of beam remnants, the thrust axis provides an estimate of the longitudinal color field between the initially created outgoing $q\overline{q}$ pair and is therefore the sensible choice in \ee collisions to search for collective effects.
Similarly to the previous figure, in order to translate the ALEPH multiplicity into the ALICE acceptance range, a scaling factor is estimated with PYTHIA~8.3 events by counting the resulting particles in the acceptance ranges of both experiments ($|\eta|<1.738$, $\pt > 0.2\,\GeVc$ in case of ALEPH). 
It is inherently difficult to compare the multiplicity in these two collision systems which have more than two orders of magnitude difference in collision energy as well as a different initial state leading to different flavor composition as well as multiplicity and momentum distributions. In order to give justice to these differences, this procedure of estimating the experimental acceptance was performed in both pp collisions at \thirteen and \ee collisions at \nineone and \alephhighenergy with resulting correction coefficients $\cpp = 0.57$, $\ceeA = 0.78$ and $\ceeB = 0.72$, respectively. 
The large difference between these two estimations reflects the different underlying mechanisms leading to multiplicity production in pp and \ee collisions and is depicted by the horizontal uncertainty bars of the ALEPH ridge yields which are given as limits at 95\% CL.
In the multiplicity range 8 to 18 (24) the yields in pp collisions are substantially above the ALEPH limit at \nineone (\alephhighenergy) while outside this range the limits from \ee collisions are above the pp measurement. The ALEPH measurement at a multiplicity of about 34 at \alephhighenergy is $1.02\sigma$ above 0. Within its large uncertainty, it is compatible with our measurement.

In order to quantify this finding the significance of the result in pp collisions to be above the one in \ee collisions is computed. The ALICE result is linearly interpolated between the two closest points to match the multiplicity. The results at different multiplicities are combined assuming the systematic uncertainty to be fully correlated across multiplicity intervals. The resulting significance of the pp measurement to be above the one in \ee collisions at \nineone is $3.8\sigma$ (using \cpp) and $5.0\sigma$ (using \ceeA). At \alephhighenergy, these significances are $5.0\sigma$ (using \cpp) and $6.3\sigma$ (using \ceeB) with the mentioned assumptions on the multiplicity conversion between the two experiments and systems.
Because of the precision of the ALEPH measurements, the multiplicity range which contributes mostly is between 8 and 24. 
In this range a near-side ridge is clearly present in pp collisions, suggesting that the processes involved in \ee annihilations do not contribute significantly to the emergence of long-range correlations in pp collisions at low multiplicity.

In Fig.~\ref{fig:ridgemodels}, the near-side yields are compared to the predictions of PYTHIA~8.3 with the Monash tune~\cite{Skands:2014pea} and the string shoving tune ($g=3$)~\cite{Bierlich:2017vhg}, as well as EPOS LHC calculations. For the model calculations, a long-range definition of $2<|\dEta|<4$ is used, as all of the models overestimate the width of the jet fragmentation peak. Under proper normalization, the choice of long-range definition does not affect the comparison, as the correlation is independent of $\dEta$~\cite{CMS:2015fgy} in this region, and the results can be directly compared.
All models are found to underestimate the data in the examined multiplicity region, although PYTHIA with shoving and EPOS LHC do exhibit collective-like signals at $N_\mathrm{ch}\gtrsim 10$ and $N_\mathrm{ch}\gtrsim 24$, respectively.
In contrast, the Monash tune as the no-ridge reference does not reproduce the near side at all, and the yield remains zero across the entire multiplicity range. Only EPOS LHC describes quantitatively the magnitude of the yield at $\langle N_\mathrm{ch}\rangle\geq 48$. These observations suggest that none of the models can fully capture the physics underlying the emergence of the near-side associated yield in low multiplicity pp collisions.

%% file: conclusions.tex
The high precision of this measurement allows one to draw quantitative comparisons between the ridge yield of a very small hadronic collision systems to the ridge yield measured in even simpler and well understood \ee annihilations. The results presented in this Letter suggest that the ridge yield measured from a hadronic system of roughly equivalent multiplicity is non-zero and substantially larger than the limit observed in \ee annihilations. Based on this, one can conclude that additional processes besides those in the \ee annihilations must play a role for the emergence of long-range correlations in pp collisions.

At the same time, the description of the ridge yields in well-established models is investigated. Calculations from three different models show that the ridge yield in the low multiplicity region in general is not reproduced. This suggests that the mechanisms for ridge yield production in very small hadronic collisions have not been understood and more theoretical work is needed.

%% file: fa_2023-11-06_Opt_C.tex

The ALICE Collaboration would like to thank all its engineers and technicians for their invaluable contributions to the construction of the experiment and the CERN accelerator teams for the outstanding performance of the LHC complex.
The ALICE Collaboration gratefully acknowledges the resources and support provided by all Grid centres and the Worldwide LHC Computing Grid (WLCG) collaboration.
The ALICE Collaboration acknowledges the following funding agencies for their support in building and running the ALICE detector:
A. I. Alikhanyan National Science Laboratory (Yerevan Physics Institute) Foundation (ANSL), State Committee of Science and World Federation of Scientists (WFS), Armenia;
Austrian Academy of Sciences, Austrian Science Fund (FWF): [M 2467-N36] and Nationalstiftung f\"{u}r Forschung, Technologie und Entwicklung, Austria;
Ministry of Communications and High Technologies, National Nuclear Research Center, Azerbaijan;
Conselho Nacional de Desenvolvimento Cient\'{\i}fico e Tecnol\'{o}gico (CNPq), Financiadora de Estudos e Projetos (Finep), Funda\c{c}\~{a}o de Amparo \`{a} Pesquisa do Estado de S\~{a}o Paulo (FAPESP) and Universidade Federal do Rio Grande do Sul (UFRGS), Brazil;
Bulgarian Ministry of Education and Science, within the National Roadmap for Research Infrastructures 2020-2027 (object CERN), Bulgaria;
Ministry of Education of China (MOEC) , Ministry of Science \& Technology of China (MSTC) and National Natural Science Foundation of China (NSFC), China;
Ministry of Science and Education and Croatian Science Foundation, Croatia;
Centro de Aplicaciones Tecnol\'{o}gicas y Desarrollo Nuclear (CEADEN), Cubaenerg\'{\i}a, Cuba;
Ministry of Education, Youth and Sports of the Czech Republic, Czech Republic;
The Danish Council for Independent Research | Natural Sciences, the VILLUM FONDEN and Danish National Research Foundation (DNRF), Denmark;
Helsinki Institute of Physics (HIP), Finland;
Commissariat \`{a} l'Energie Atomique (CEA) and Institut National de Physique Nucl\'{e}aire et de Physique des Particules (IN2P3) and Centre National de la Recherche Scientifique (CNRS), France;
Bundesministerium f\"{u}r Bildung und Forschung (BMBF) and GSI Helmholtzzentrum f\"{u}r Schwerionenforschung GmbH, Germany;
General Secretariat for Research and Technology, Ministry of Education, Research and Religions, Greece;
National Research, Development and Innovation Office, Hungary;
Department of Atomic Energy Government of India (DAE), Department of Science and Technology, Government of India (DST), University Grants Commission, Government of India (UGC) and Council of Scientific and Industrial Research (CSIR), India;
National Research and Innovation Agency - BRIN, Indonesia;
Istituto Nazionale di Fisica Nucleare (INFN), Italy;
Japanese Ministry of Education, Culture, Sports, Science and Technology (MEXT) and Japan Society for the Promotion of Science (JSPS) KAKENHI, Japan;
Consejo Nacional de Ciencia (CONACYT) y Tecnolog\'{i}a, through Fondo de Cooperaci\'{o}n Internacional en Ciencia y Tecnolog\'{i}a (FONCICYT) and Direcci\'{o}n General de Asuntos del Personal Academico (DGAPA), Mexico;
Nederlandse Organisatie voor Wetenschappelijk Onderzoek (NWO), Netherlands;
The Research Council of Norway, Norway;
Commission on Science and Technology for Sustainable Development in the South (COMSATS), Pakistan;
Pontificia Universidad Cat\'{o}lica del Per\'{u}, Peru;
Ministry of Education and Science, National Science Centre and WUT ID-UB, Poland;
Korea Institute of Science and Technology Information and National Research Foundation of Korea (NRF), Republic of Korea;
Ministry of Education and Scientific Research, Institute of Atomic Physics, Ministry of Research and Innovation and Institute of Atomic Physics and Universitatea Nationala de Stiinta si Tehnologie Politehnica Bucuresti, Romania;
Ministry of Education, Science, Research and Sport of the Slovak Republic, Slovakia;
National Research Foundation of South Africa, South Africa;
Swedish Research Council (VR) and Knut \& Alice Wallenberg Foundation (KAW), Sweden;
European Organization for Nuclear Research, Switzerland;
Suranaree University of Technology (SUT), National Science and Technology Development Agency (NSTDA) and National Science, Research and Innovation Fund (NSRF via PMU-B B05F650021), Thailand;
National Academy of  Sciences of Ukraine, Ukraine;
Science and Technology Facilities Council (STFC), United Kingdom;
National Science Foundation of the United States of America (NSF) and United States Department of Energy, Office of Nuclear Physics (DOE NP), United States of America.
In addition, individual groups or members have received support from:
European Research Council, Strong 2020 - Horizon 2020 (grant nos. 950692, 824093), European Union;
Academy of Finland (Center of Excellence in Quark Matter) (grant nos. 346327, 346328), Finland.

%% file: 2023-11-06-Alice_Authorlist_2023-11-06_Opt_C.tex
\begin{flushleft} 
\small

S.~Acharya\,\orcidlink{0000-0002-9213-5329}\,$^{\rm 127}$, 
D.~Adamov\'{a}\,\orcidlink{0000-0002-0504-7428}\,$^{\rm 86}$, 
G.~Aglieri Rinella\,\orcidlink{0000-0002-9611-3696}\,$^{\rm 33}$, 
L.~Aglietta$^{\rm 25}$, 
M.~Agnello\,\orcidlink{0000-0002-0760-5075}\,$^{\rm 30}$, 
N.~Agrawal\,\orcidlink{0000-0003-0348-9836}\,$^{\rm 52}$, 
Z.~Ahammed\,\orcidlink{0000-0001-5241-7412}\,$^{\rm 135}$, 
S.~Ahmad\,\orcidlink{0000-0003-0497-5705}\,$^{\rm 16}$, 
S.U.~Ahn\,\orcidlink{0000-0001-8847-489X}\,$^{\rm 72}$, 
I.~Ahuja\,\orcidlink{0000-0002-4417-1392}\,$^{\rm 38}$
% ...

\section*{Affiliation Notes}

$^{\rm I}$ Deceased\\
$^{\rm II}$ Also at: Max-Planck-Institut fur Physik, Munich, Germany\\
$^{\rm III}$ Also at: Italian National Agency for New Technologies, Energy and Sustainable Economic Development (ENEA), Bologna, Italy\\
$^{\rm IV}$ Also at: Dipartimento DET del Politecnico di Torino, Turin, Italy\\
$^{\rm V}$ Also at: Department of Applied Physics, Aligarh Muslim University, Aligarh, India\\
$^{\rm VI}$ Also at: Institute of Theoretical Physics, University of Wroclaw, Poland\\
$^{\rm VII}$ Also at: An institution covered by a cooperation agreement with CERN\\

\section*{Collaboration Institutes}

$^{1}$ A.I. Alikhanyan National Science Laboratory (Yerevan Physics Institute) Foundation, Yerevan, Armenia\\
$^{2}$ AGH University of Krakow, Cracow, Poland\\
$^{3}$ Bogolyubov Institute for Theoretical Physics, National Academy of Sciences of Ukraine, Kiev, Ukraine\\
$^{4}$ Bose Institute, Department of Physics  and Centre for Astroparticle Physics and Space Science (CAPSS), Kolkata, India\\
$^{5}$ California Polytechnic State University, San Luis Obispo, California, United States\\
$^{6}$ Central China Normal University, Wuhan, China\\
$^{7}$ Centro de Aplicaciones Tecnol\'{o}gicas y Desarrollo Nuclear (CEADEN), Havana, Cuba\\
$^{8}$ Centro de Investigaci\'{o}n y de Estudios Avanzados (CINVESTAV), Mexico City and M\'{e}rida, Mexico\\
$^{9}$ Chicago State University, Chicago, Illinois, United States\\
$^{10}$ China Institute of Atomic Energy, Beijing, China\\
$^{11}$ China University of Geosciences, Wuhan, China\\
$^{12}$ Chungbuk National University, Cheongju, Republic of Korea\\
$^{13}$ Comenius University Bratislava, Faculty of Mathematics, Physics and Informatics, Bratislava, Slovak Republic\\
$^{14}$ COMSATS University Islamabad, Islamabad, Pakistan\\
$^{15}$ Creighton University, Omaha, Nebraska, United States\\
$^{16}$ Department of Physics, Aligarh Muslim University, Aligarh, India\\
$^{17}$ Department of Physics, Pusan National University, Pusan, Republic of Korea\\
$^{18}$ Department of Physics, Sejong University, Seoul, Republic of Korea\\
$^{19}$ Department of Physics, University of California, Berkeley, California, United States\\
$^{20}$ Department of Physics, University of Oslo, Oslo, Norway\\
$^{21}$ Department of Physics and Technology, University of Bergen, Bergen, Norway\\
$^{22}$ Dipartimento di Fisica, Universit\`{a} di Pavia, Pavia, Italy\\
$^{23}$ Dipartimento di Fisica dell'Universit\`{a} and Sezione INFN, Cagliari, Italy\\
$^{24}$ Dipartimento di Fisica dell'Universit\`{a} and Sezione INFN, Trieste, Italy\\
$^{25}$ Dipartimento di Fisica dell'Universit\`{a} and Sezione INFN, Turin, Italy\\
$^{26}$ Dipartimento di Fisica e Astronomia dell'Universit\`{a} and Sezione INFN, Bologna, Italy\\
$^{27}$ Dipartimento di Fisica e Astronomia dell'Universit\`{a} and Sezione INFN, Catania, Italy\\
$^{28}$ Dipartimento di Fisica e Astronomia dell'Universit\`{a} and Sezione INFN, Padova, Italy\\
$^{29}$ Dipartimento di Fisica `E.R.~Caianiello' dell'Universit\`{a} and Gruppo Collegato INFN, Salerno, Italy\\
$^{30}$ Dipartimento DISAT del Politecnico and Sezione INFN, Turin, Italy\\
$^{31}$ Dipartimento di Scienze MIFT, Universit\`{a} di Messina, Messina, Italy\\
$^{32}$ Dipartimento Interateneo di Fisica `M.~Merlin' and Sezione INFN, Bari, Italy\\
$^{33}$ European Organization for Nuclear Research (CERN), Geneva, Switzerland\\
$^{34}$ Faculty of Electrical Engineering, Mechanical Engineering and Naval Architecture, University of Split, Split, Croatia\\
$^{35}$ Faculty of Engineering and Science, Western Norway University of Applied Sciences, Bergen, Norway\\
$^{36}$ Faculty of Nuclear Sciences and Physical Engineering, Czech Technical University in Prague, Prague, Czech Republic\\
$^{37}$ Faculty of Physics, Sofia University, Sofia, Bulgaria\\
$^{38}$ Faculty of Science, P.J.~\v{S}af\'{a}rik University, Ko\v{s}ice, Slovak Republic\\
$^{39}$ Frankfurt Institute for Advanced Studies, Johann Wolfgang Goethe-Universit\"{a}t Frankfurt, Frankfurt, Germany\\
$^{40}$ Fudan University, Shanghai, China\\
$^{41}$ Gangneung-Wonju National University, Gangneung, Republic of Korea\\
$^{42}$ Gauhati University, Department of Physics, Guwahati, India\\
$^{43}$ Helmholtz-Institut f\"{u}r Strahlen- und Kernphysik, Rheinische Friedrich-Wilhelms-Universit\"{a}t Bonn, Bonn, Germany\\
$^{44}$ Helsinki Institute of Physics (HIP), Helsinki, Finland\\
$^{45}$ High Energy Physics Group,  Universidad Aut\'{o}noma de Puebla, Puebla, Mexico\\
$^{46}$ Horia Hulubei National Institute of Physics and Nuclear Engineering, Bucharest, Romania\\
$^{47}$ HUN-REN Wigner Research Centre for Physics, Budapest, Hungary\\
$^{48}$ Indian Institute of Technology Bombay (IIT), Mumbai, India\\
$^{49}$ Indian Institute of Technology Indore, Indore, India\\
$^{50}$ INFN, Laboratori Nazionali di Frascati, Frascati, Italy\\
$^{51}$ INFN, Sezione di Bari, Bari, Italy\\
$^{52}$ INFN, Sezione di Bologna, Bologna, Italy\\
$^{53}$ INFN, Sezione di Cagliari, Cagliari, Italy\\
$^{54}$ INFN, Sezione di Catania, Catania, Italy\\
$^{55}$ INFN, Sezione di Padova, Padova, Italy\\
$^{56}$ INFN, Sezione di Pavia, Pavia, Italy\\
$^{57}$ INFN, Sezione di Torino, Turin, Italy\\
$^{58}$ INFN, Sezione di Trieste, Trieste, Italy\\
$^{59}$ Inha University, Incheon, Republic of Korea\\
$^{60}$ Institute for Gravitational and Subatomic Physics (GRASP), Utrecht University/Nikhef, Utrecht, Netherlands\\
$^{61}$ Institute of Experimental Physics, Slovak Academy of Sciences, Ko\v{s}ice, Slovak Republic\\
$^{62}$ Institute of Physics, Homi Bhabha National Institute, Bhubaneswar, India\\
$^{63}$ Institute of Physics of the Czech Academy of Sciences, Prague, Czech Republic\\
$^{64}$ Institute of Space Science (ISS), Bucharest, Romania\\
$^{65}$ Institut f\"{u}r Kernphysik, Johann Wolfgang Goethe-Universit\"{a}t Frankfurt, Frankfurt, Germany\\
$^{66}$ Instituto de Ciencias Nucleares, Universidad Nacional Aut\'{o}noma de M\'{e}xico, Mexico City, Mexico\\
$^{67}$ Instituto de F\'{i}sica, Universidade Federal do Rio Grande do Sul (UFRGS), Porto Alegre, Brazil\\
$^{68}$ Instituto de F\'{\i}sica, Universidad Nacional Aut\'{o}noma de M\'{e}xico, Mexico City, Mexico\\
$^{69}$ iThemba LABS, National Research Foundation, Somerset West, South Africa\\
$^{70}$ Jeonbuk National University, Jeonju, Republic of Korea\\
$^{71}$ Johann-Wolfgang-Goethe Universit\"{a}t Frankfurt Institut f\"{u}r Informatik, Fachbereich Informatik und Mathematik, Frankfurt, Germany\\
$^{72}$ Korea Institute of Science and Technology Information, Daejeon, Republic of Korea\\
$^{73}$ Laboratoire de Physique Subatomique et de Cosmologie, Universit\'{e} Grenoble-Alpes, CNRS-IN2P3, Grenoble, France\\
$^{74}$ Lawrence Berkeley National Laboratory, Berkeley, California, United States\\
$^{75}$ Lund University Department of Physics, Division of Particle Physics, Lund, Sweden\\
$^{76}$ Nagasaki Institute of Applied Science, Nagasaki, Japan\\
$^{77}$ Nara Women{'}s University (NWU), Nara, Japan\\
$^{78}$ National and Kapodistrian University of Athens, School of Science, Department of Physics , Athens, Greece\\
$^{79}$ National Centre for Nuclear Research, Warsaw, Poland\\
$^{80}$ National Institute of Science Education and Research, Homi Bhabha National Institute, Jatni, India\\
$^{81}$ National Nuclear Research Center, Baku, Azerbaijan\\
$^{82}$ National Research and Innovation Agency - BRIN, Jakarta, Indonesia\\
$^{83}$ Niels Bohr Institute, University of Copenhagen, Copenhagen, Denmark\\
$^{84}$ Nikhef, National institute for subatomic physics, Amsterdam, Netherlands\\
$^{85}$ Nuclear Physics Group, STFC Daresbury Laboratory, Daresbury, United Kingdom\\
$^{86}$ Nuclear Physics Institute of the Czech Academy of Sciences, Husinec-\v{R}e\v{z}, Czech Republic\\
$^{87}$ Oak Ridge National Laboratory, Oak Ridge, Tennessee, United States\\
$^{88}$ Ohio State University, Columbus, Ohio, United States\\
$^{89}$ Physics department, Faculty of science, University of Zagreb, Zagreb, Croatia\\
$^{90}$ Physics Department, Panjab University, Chandigarh, India\\
$^{91}$ Physics Department, University of Jammu, Jammu, India\\
$^{92}$ Physics Program and International Institute for Sustainability with Knotted Chiral Meta Matter (SKCM2), Hiroshima University, Hiroshima, Japan\\
$^{93}$ Physikalisches Institut, Eberhard-Karls-Universit\"{a}t T\"{u}bingen, T\"{u}bingen, Germany\\
$^{94}$ Physikalisches Institut, Ruprecht-Karls-Universit\"{a}t Heidelberg, Heidelberg, Germany\\
$^{95}$ Physik Department, Technische Universit\"{a}t M\"{u}nchen, Munich, Germany\\
$^{96}$ Politecnico di Bari and Sezione INFN, Bari, Italy\\
$^{97}$ Research Division and ExtreMe Matter Institute EMMI, GSI Helmholtzzentrum f\"ur Schwerionenforschung GmbH, Darmstadt, Germany\\
$^{98}$ Saga University, Saga, Japan\\
$^{99}$ Saha Institute of Nuclear Physics, Homi Bhabha National Institute, Kolkata, India\\
$^{100}$ School of Physics and Astronomy, University of Birmingham, Birmingham, United Kingdom\\
$^{101}$ Secci\'{o}n F\'{\i}sica, Departamento de Ciencias, Pontificia Universidad Cat\'{o}lica del Per\'{u}, Lima, Peru\\
$^{102}$ Stefan Meyer Institut f\"{u}r Subatomare Physik (SMI), Vienna, Austria\\
$^{103}$ SUBATECH, IMT Atlantique, Nantes Universit\'{e}, CNRS-IN2P3, Nantes, France\\
$^{104}$ Sungkyunkwan University, Suwon City, Republic of Korea\\
$^{105}$ Suranaree University of Technology, Nakhon Ratchasima, Thailand\\
$^{106}$ Technical University of Ko\v{s}ice, Ko\v{s}ice, Slovak Republic\\
$^{107}$ The Henryk Niewodniczanski Institute of Nuclear Physics, Polish Academy of Sciences, Cracow, Poland\\
$^{108}$ The University of Texas at Austin, Austin, Texas, United States\\
$^{109}$ Universidad Aut\'{o}noma de Sinaloa, Culiac\'{a}n, Mexico\\
$^{110}$ Universidade de S\~{a}o Paulo (USP), S\~{a}o Paulo, Brazil\\
$^{111}$ Universidade Estadual de Campinas (UNICAMP), Campinas, Brazil\\
$^{112}$ Universidade Federal do ABC, Santo Andre, Brazil\\
$^{113}$ Universitatea Nationala de Stiinta si Tehnologie Politehnica Bucuresti, Bucharest, Romania\\
$^{114}$ University of Cape Town, Cape Town, South Africa\\
$^{115}$ University of Derby, Derby, United Kingdom\\
$^{116}$ University of Houston, Houston, Texas, United States\\
$^{117}$ University of Jyv\"{a}skyl\"{a}, Jyv\"{a}skyl\"{a}, Finland\\
$^{118}$ University of Kansas, Lawrence, Kansas, United States\\
$^{119}$ University of Liverpool, Liverpool, United Kingdom\\
$^{120}$ University of Science and Technology of China, Hefei, China\\
$^{121}$ University of South-Eastern Norway, Kongsberg, Norway\\
$^{122}$ University of Tennessee, Knoxville, Tennessee, United States\\
$^{123}$ University of the Witwatersrand, Johannesburg, South Africa\\
$^{124}$ University of Tokyo, Tokyo, Japan\\
$^{125}$ University of Tsukuba, Tsukuba, Japan\\
$^{126}$ Universit\"{a}t M\"{u}nster, Institut f\"{u}r Kernphysik, M\"{u}nster, Germany\\
$^{127}$ Universit\'{e} Clermont Auvergne, CNRS/IN2P3, LPC, Clermont-Ferrand, France\\
$^{128}$ Universit\'{e} de Lyon, CNRS/IN2P3, Institut de Physique des 2 Infinis de Lyon, Lyon, France\\
$^{129}$ Universit\'{e} de Strasbourg, CNRS, IPHC UMR 7178, F-67000 Strasbourg, France, Strasbourg, France\\
$^{130}$ Universit\'{e} Paris-Saclay, Centre d'Etudes de Saclay (CEA), IRFU, D\'{e}partment de Physique Nucl\'{e}aire (DPhN), Saclay, France\\
$^{131}$ Universit\'{e}  Paris-Saclay, CNRS/IN2P3, IJCLab, Orsay, France\\
$^{132}$ Universit\`{a} degli Studi di Foggia, Foggia, Italy\\
$^{133}$ Universit\`{a} del Piemonte Orientale, Vercelli, Italy\\
$^{134}$ Universit\`{a} di Brescia, Brescia, Italy\\
$^{135}$ Variable Energy Cyclotron Centre, Homi Bhabha National Institute, Kolkata, India\\
$^{136}$ Warsaw University of Technology, Warsaw, Poland\\
$^{137}$ Wayne State University, Detroit, Michigan, United States\\
$^{138}$ Yale University, New Haven, Connecticut, United States\\
$^{139}$ Yonsei University, Seoul, Republic of Korea\\
$^{140}$  Zentrum  f\"{u}r Technologie und Transfer (ZTT), Worms, Germany\\
$^{141}$ Affiliated with an institute covered by a cooperation agreement with CERN\\
$^{142}$ Affiliated with an international laboratory covered by a cooperation agreement with CERN.\\

\end{flushleft}

%% file: main.bbl
\providecommand{\href}[2]{#2}\begingroup\raggedright\begin{thebibliography}{10}

\bibitem{Adams:2005dq}
{\bfseries STAR} Collaboration, J.~Adams {\em et~al.}, ``{Experimental and
  theoretical challenges in the search for the quark gluon plasma: The STAR
  Collaboration's critical assessment of the evidence from RHIC collisions}'',
  \href{http://dx.doi.org/10.1016/j.nuclphysa.2005.03.085}{{\em Nucl. Phys.}
  {\bfseries A757} (2005) 102--183},
\href{http://arxiv.org/abs/nucl-ex/0501009}{{\ttfamily arXiv:nucl-ex/0501009
  [nucl-ex]}}.

\bibitem{Adcox:2004mh}
{\bfseries PHENIX} Collaboration, K.~Adcox {\em et~al.}, ``{Formation of dense
  partonic matter in relativistic nucleus-nucleus collisions at RHIC:
  Experimental evaluation by the PHENIX collaboration}'',
  \href{http://dx.doi.org/10.1016/j.nuclphysa.2005.03.086}{{\em Nucl. Phys.}
  {\bfseries A757} (2005) 184--283},
\href{http://arxiv.org/abs/nucl-ex/0410003}{{\ttfamily arXiv:nucl-ex/0410003
  [nucl-ex]}}.

\bibitem{Arsene:2004fa}
{\bfseries BRAHMS} Collaboration, I.~Arsene {\em et~al.}, ``{Quark gluon plasma
  and color glass condensate at RHIC? The Perspective from the BRAHMS
  experiment}'', \href{http://dx.doi.org/10.1016/j.nuclphysa.2005.02.130}{{\em
  Nucl. Phys.} {\bfseries A757} (2005) 1--27},
\href{http://arxiv.org/abs/nucl-ex/0410020}{{\ttfamily arXiv:nucl-ex/0410020
  [nucl-ex]}}.

\bibitem{Back:2004je}
{\bfseries PHOBOS} Collaboration, B.~B. Back {\em et~al.}, ``{The PHOBOS
  perspective on discoveries at RHIC}'',
  \href{http://dx.doi.org/10.1016/j.nuclphysa.2005.03.084}{{\em Nucl. Phys.}
  {\bfseries A757} (2005) 28--101},
\href{http://arxiv.org/abs/nucl-ex/0410022}{{\ttfamily arXiv:nucl-ex/0410022
  [nucl-ex]}}.

\bibitem{Abelev:2012di}
{\bfseries ALICE} Collaboration, B.~Abelev {\em et~al.}, ``{Anisotropic flow of
  charged hadrons, pions and (anti-)protons measured at high transverse
  momentum in Pb--Pb collisions at $\sqrt{s_\mathrm{NN}}$=2.76 TeV}'',
  \href{http://dx.doi.org/10.1016/j.physletb.2012.12.066}{{\em Phys. Lett.}
  {\bfseries B719} (2013) 18--28},
\href{http://arxiv.org/abs/1205.5761}{{\ttfamily arXiv:1205.5761 [nucl-ex]}}.

\bibitem{Abelev:2014pua}
{\bfseries ALICE} Collaboration, B.~Abelev {\em et~al.}, ``{Elliptic flow of
  identified hadrons in Pb--Pb collisions at $ \sqrt{s_{\mathrm{NN}}}=2.76 $
  TeV}'', \href{http://dx.doi.org/10.1007/JHEP06(2015)190}{{\em JHEP}
  {\bfseries 06} (2015) 190},
\href{http://arxiv.org/abs/1405.4632}{{\ttfamily arXiv:1405.4632 [nucl-ex]}}.

\bibitem{ATLAS:2011ah}
{\bfseries ATLAS} Collaboration, G.~Aad {\em et~al.}, ``{Measurement of the
  pseudorapidity and transverse momentum dependence of the elliptic flow of
  charged particles in lead-lead collisions at $\sqrt{s_\mathrm{NN}}=2.76$ TeV
  with the ATLAS detector}'',
  \href{http://dx.doi.org/10.1016/j.physletb.2011.12.056}{{\em Phys. Lett. B}
  {\bfseries 707} (2012) 330--348},
  \href{http://arxiv.org/abs/1108.6018}{{\ttfamily arXiv:1108.6018 [hep-ex]}}.

\bibitem{CMS:2010ifv}
{\bfseries CMS} Collaboration, V.~Khachatryan {\em et~al.}, ``{Observation of
  Long-Range Near-Side Angular Correlations in Proton-Proton Collisions at the
  LHC}'', \href{http://dx.doi.org/10.1007/JHEP09(2010)091}{{\em JHEP}
  {\bfseries 09} (2010) 091}, \href{http://arxiv.org/abs/1009.4122}{{\ttfamily
  arXiv:1009.4122 [hep-ex]}}.

\bibitem{ATLAS:2015hzw}
{\bfseries ATLAS} Collaboration, G.~Aad {\em et~al.}, ``{Observation of
  Long-Range Elliptic Azimuthal Anisotropies in $\sqrt{s}=$13 and 2.76 TeV $pp$
  Collisions with the ATLAS Detector}'',
  \href{http://dx.doi.org/10.1103/PhysRevLett.116.172301}{{\em Phys. Rev.
  Lett.} {\bfseries 116} (2016) 172301},
  \href{http://arxiv.org/abs/1509.04776}{{\ttfamily arXiv:1509.04776
  [hep-ex]}}.

\bibitem{Khachatryan:2015lva}
{\bfseries CMS} Collaboration, V.~Khachatryan {\em et~al.}, ``{Measurement of
  long-range near-side two-particle angular correlations in pp collisions at
  $\sqrt s =$13 TeV}'',
  \href{http://dx.doi.org/10.1103/PhysRevLett.116.172302}{{\em Phys. Rev.
  Lett.} {\bfseries 116} (2016) 172302},
  \href{http://arxiv.org/abs/1510.03068}{{\ttfamily arXiv:1510.03068
  [nucl-ex]}}.

\bibitem{Khachatryan:2016txc}
{\bfseries CMS} Collaboration, V.~Khachatryan {\em et~al.}, ``{Evidence for
  collectivity in pp collisions at the LHC}'',
  \href{http://dx.doi.org/10.1016/j.physletb.2016.12.009}{{\em Phys. Lett. B}
  {\bfseries 765} (2017) 193--220},
  \href{http://arxiv.org/abs/1606.06198}{{\ttfamily arXiv:1606.06198
  [nucl-ex]}}.

\bibitem{Acharya:2019vdf}
{\bfseries ALICE} Collaboration, S.~Acharya {\em et~al.}, ``{Investigations of
  Anisotropic Flow Using Multiparticle Azimuthal Correlations in pp, p--Pb,
  Xe--Xe, and Pb--Pb Collisions at the LHC}'',
  \href{http://dx.doi.org/10.1103/PhysRevLett.123.142301}{{\em Phys. Rev.
  Lett.} {\bfseries 123} (2019) 142301},
  \href{http://arxiv.org/abs/1903.01790}{{\ttfamily arXiv:1903.01790
  [nucl-ex]}}.

\bibitem{ATLAS:2017rtr}
{\bfseries ATLAS} Collaboration, M.~Aaboud {\em et~al.}, ``{Measurement of
  long-range multiparticle azimuthal correlations with the subevent cumulant
  method in $pp$ and $p + Pb$ collisions with the ATLAS detector at the CERN
  Large Hadron Collider}'',
  \href{http://dx.doi.org/10.1103/PhysRevC.97.024904}{{\em Phys. Rev. C}
  {\bfseries 97} (2018) 024904},
  \href{http://arxiv.org/abs/1708.03559}{{\ttfamily arXiv:1708.03559
  [hep-ex]}}.

\bibitem{ALICE:2012eyl}
{\bfseries ALICE} Collaboration, B.~Abelev {\em et~al.}, ``{Long-range angular
  correlations on the near and away side in p--Pb collisions at
  $\sqrt{s_\mathrm{NN}}=5.02$ TeV}'',
  \href{http://dx.doi.org/10.1016/j.physletb.2013.01.012}{{\em Phys. Lett. B}
  {\bfseries 719} (2013) 29--41},
  \href{http://arxiv.org/abs/1212.2001}{{\ttfamily arXiv:1212.2001 [nucl-ex]}}.

\bibitem{ATLAS:2014qaj}
{\bfseries ATLAS} Collaboration, G.~Aad {\em et~al.}, ``{Measurement of
  long-range pseudorapidity correlations and azimuthal harmonics in
  $\sqrt{s_\mathrm{NN}}=5.02$ TeV proton-lead collisions with the ATLAS
  detector}'', \href{http://dx.doi.org/10.1103/PhysRevC.90.044906}{{\em Phys.
  Rev. C} {\bfseries 90} (2014) 044906},
  \href{http://arxiv.org/abs/1409.1792}{{\ttfamily arXiv:1409.1792 [hep-ex]}}.

\bibitem{ATLAS:2016yzd}
{\bfseries ATLAS} Collaboration, M.~Aaboud {\em et~al.}, ``{Measurements of
  long-range azimuthal anisotropies and associated Fourier coefficients for
  $pp$ collisions at $\sqrt{s}=5.02$ and $13$ TeV and $p$+Pb collisions at
  $\sqrt{s_{\mathrm{NN}}}=5.02$ TeV with the ATLAS detector}'',
  \href{http://dx.doi.org/10.1103/PhysRevC.96.024908}{{\em Phys. Rev. C}
  {\bfseries 96} (2017) 024908},
  \href{http://arxiv.org/abs/1609.06213}{{\ttfamily arXiv:1609.06213
  [nucl-ex]}}.

\bibitem{Khachatryan:2016ibd}
{\bfseries CMS} Collaboration, V.~Khachatryan {\em et~al.}, ``{Pseudorapidity
  dependence of long-range two-particle correlations in $p$Pb collisions at
  $\sqrt {s_\mathrm{NN}}=$ 5.02 TeV}'',
  \href{http://dx.doi.org/10.1103/PhysRevC.96.014915}{{\em Phys. Rev. C}
  {\bfseries 96} (2017) 014915},
  \href{http://arxiv.org/abs/1604.05347}{{\ttfamily arXiv:1604.05347
  [nucl-ex]}}.

\bibitem{PHENIX:2018lia}
{\bfseries PHENIX} Collaboration, C.~Aidala {\em et~al.}, ``{Creation of
  quark--gluon plasma droplets with three distinct geometries}'',
  \href{http://dx.doi.org/10.1038/s41567-018-0360-0}{{\em Nature Phys.}
  {\bfseries 15} (2019) 214--220},
  \href{http://arxiv.org/abs/1805.02973}{{\ttfamily arXiv:1805.02973
  [nucl-ex]}}.

\bibitem{Aidala:2017ajz}
{\bfseries PHENIX} Collaboration, C.~Aidala {\em et~al.}, ``{Measurements of
  Multiparticle Correlations in $d+\mathrm{Au}$ Collisions at 200, 62.4, 39,
  and 19.6 GeV and $p+\mathrm{Au}$ Collisions at 200 GeV and Implications for
  Collective Behavior}'',
  \href{http://dx.doi.org/10.1103/PhysRevLett.120.062302}{{\em Phys. Rev.
  Lett.} {\bfseries 120} (2018) 062302},
  \href{http://arxiv.org/abs/1707.06108}{{\ttfamily arXiv:1707.06108
  [nucl-ex]}}.

\bibitem{STAR:2015kak}
{\bfseries STAR} Collaboration, L.~Adamczyk {\em et~al.}, ``{Long-range
  pseudorapidity dihadron correlations in $d$+Au collisions at $\sqrt{s_{\rm
  NN}}=200$ GeV}'',
  \href{http://dx.doi.org/10.1016/j.physletb.2015.05.075}{{\em Phys. Lett. B}
  {\bfseries 747} (2015) 265--271},
  \href{http://arxiv.org/abs/1502.07652}{{\ttfamily arXiv:1502.07652
  [nucl-ex]}}.

\bibitem{Voloshin:2008dg}
S.~A. Voloshin, A.~M. Poskanzer, and R.~Snellings, ``{Collective phenomena in
  non-central nuclear collisions}'',
  \href{http://dx.doi.org/10.1007/978-3-642-01539-7_10}{{\em Landolt-Bornstein}
  {\bfseries 23} (2010) 293--333},
  \href{http://arxiv.org/abs/0809.2949}{{\ttfamily arXiv:0809.2949 [nucl-ex]}}.

\bibitem{Nagle:2018nvi}
J.~L. Nagle and W.~A. Zajc, ``{Small System Collectivity in Relativistic
  Hadronic and Nuclear Collisions}'',
  \href{http://dx.doi.org/10.1146/annurev-nucl-101916-123209}{{\em Ann. Rev.
  Nucl. Part. Sci.} {\bfseries 68} (2018) 211--235},
  \href{http://arxiv.org/abs/1801.03477}{{\ttfamily arXiv:1801.03477
  [nucl-ex]}}.

\bibitem{Strickland:2018exs}
M.~Strickland, ``{Small system studies: A theory overview}'',
  \href{http://dx.doi.org/10.1016/j.nuclphysa.2018.09.071}{{\em Nucl. Phys. A}
  {\bfseries 982} (2019) 92--98},
  \href{http://arxiv.org/abs/1807.07191}{{\ttfamily arXiv:1807.07191
  [nucl-th]}}.

\bibitem{Loizides:2016tew}
C.~Loizides, ``{Experimental overview on small collision systems at the LHC}'',
  \href{http://dx.doi.org/10.1016/j.nuclphysa.2016.04.022}{{\em Nucl. Phys. A}
  {\bfseries 956} (2016) 200--207},
  \href{http://arxiv.org/abs/1602.09138}{{\ttfamily arXiv:1602.09138
  [nucl-ex]}}.

\bibitem{Schenke:2020mbo}
B.~Schenke, C.~Shen, and P.~Tribedy, ``{Running the gamut of high energy
  nuclear collisions}'',
  \href{http://dx.doi.org/10.1103/PhysRevC.102.044905}{{\em Phys. Rev. C}
  {\bfseries 102} (2020) 044905},
  \href{http://arxiv.org/abs/2005.14682}{{\ttfamily arXiv:2005.14682
  [nucl-th]}}.

\bibitem{Zhao:2022ugy}
W.~Zhao, S.~Ryu, C.~Shen, and B.~Schenke, ``{3D structure of anisotropic flow
  in small collision systems at energies available at the BNL Relativistic
  Heavy Ion Collider}'',
  \href{http://dx.doi.org/10.1103/PhysRevC.107.014904}{{\em Phys. Rev. C}
  {\bfseries 107} (2023) 014904},
  \href{http://arxiv.org/abs/2211.16376}{{\ttfamily arXiv:2211.16376
  [nucl-th]}}.

\bibitem{STAR:2022pfn}
{\bfseries STAR} Collaboration, M.~I. Abdulhamid {\em et~al.}, ``{Measurements
  of the Elliptic and Triangular Azimuthal Anisotropies in Central He3+Au, d+Au
  and p+Au Collisions at $\sqrt{s_\mathrm{NN}}$ = 200\,\,GeV}'',
  \href{http://dx.doi.org/10.1103/PhysRevLett.130.242301}{{\em Phys. Rev.
  Lett.} {\bfseries 130} (2023) 242301},
  \href{http://arxiv.org/abs/2210.11352}{{\ttfamily arXiv:2210.11352
  [nucl-ex]}}.

\bibitem{Dusling:2012cg}
K.~Dusling and R.~Venugopalan, ``{Evidence for BFKL and saturation dynamics
  from dihadron spectra at the LHC}'',
  \href{http://dx.doi.org/10.1103/PhysRevD.87.051502}{{\em Phys. Rev. D}
  {\bfseries 87} (2013) 051502},
  \href{http://arxiv.org/abs/1210.3890}{{\ttfamily arXiv:1210.3890 [hep-ph]}}.

\bibitem{Bzdak:2013zma}
A.~Bzdak, B.~Schenke, P.~Tribedy, and R.~Venugopalan, ``{Initial state geometry
  and the role of hydrodynamics in proton-proton, proton-nucleus and
  deuteron-nucleus collisions}'',
  \href{http://dx.doi.org/10.1103/PhysRevC.87.064906}{{\em Phys. Rev. C}
  {\bfseries 87} (2013) 064906},
  \href{http://arxiv.org/abs/1304.3403}{{\ttfamily arXiv:1304.3403 [nucl-th]}}.

\bibitem{Greif:2017bnr}
M.~Greif, C.~Greiner, B.~Schenke, S.~Schlichting, and Z.~Xu, ``{Importance of
  initial and final state effects for azimuthal correlations in p+Pb
  collisions}'', \href{http://dx.doi.org/10.1103/PhysRevD.96.091504}{{\em Phys.
  Rev. D} {\bfseries 96} (2017) 091504},
  \href{http://arxiv.org/abs/1708.02076}{{\ttfamily arXiv:1708.02076
  [hep-ph]}}.

\bibitem{Mantysaari:2017cni}
H.~Mantysaari, B.~Schenke, C.~Shen, and P.~Tribedy, ``{Imprints of fluctuating
  proton shapes on flow in proton-lead collisions at the LHC}'',
  \href{http://dx.doi.org/10.1016/j.physletb.2017.07.038}{{\em Phys. Lett. B}
  {\bfseries 772} (2017) 681--686},
  \href{http://arxiv.org/abs/1705.03177}{{\ttfamily arXiv:1705.03177
  [nucl-th]}}.

\bibitem{Schenke:2022mjv}
B.~Schenke, S.~Schlichting, and P.~Singh, ``{Rapidity dependence of initial
  state geometry and momentum correlations in p+Pb collisions}'',
  \href{http://dx.doi.org/10.1103/PhysRevD.105.094023}{{\em Phys. Rev. D}
  {\bfseries 105} (2022) 094023},
  \href{http://arxiv.org/abs/2201.08864}{{\ttfamily arXiv:2201.08864
  [nucl-th]}}.

\bibitem{DECAMP1990121}
{\bfseries ALEPH} Collaboration, D.~Decamp {\em et~al.}, ``Aleph: A detector
  for electron-positron annihilations at lep'',
  \href{http://dx.doi.org/https://doi.org/10.1016/0168-9002(90)91831-U}{{\em
  Nuclear Instruments and Methods in Physics Research Section A: Accelerators,
  Spectrometers, Detectors and Associated Equipment} {\bfseries 294} (1990)
  121--178}.
  \url{https://www.sciencedirect.com/science/article/pii/016890029091831U}.

\bibitem{Badea:2019vey}
A.~Badea {\em et~al.}, ``{Measurements of two-particle correlations in $e^+e^-$
  collisions at 91 GeV with ALEPH archived data}'',
  \href{http://dx.doi.org/10.1103/PhysRevLett.123.212002}{{\em Phys. Rev.
  Lett.} {\bfseries 123} (2019) 212002},
  \href{http://arxiv.org/abs/1906.00489}{{\ttfamily arXiv:1906.00489
  [hep-ex]}}.

\bibitem{Chen:2023njr}
Y.-C. Chen {\em et~al.}, ``{Long-range near-side correlation in $e^+e^-$
  Collisions at 183-209 GeV with ALEPH Archived Data}'',
  \href{http://arxiv.org/abs/2312.05084}{{\ttfamily arXiv:2312.05084
  [hep-ex]}}.

\bibitem{PhysRevC.97.024909}
J.~L. Nagle, R.~Belmont, K.~Hill, J.~O. Koop, D.~V. Perepelitsa, P.~Yin, Z.-W.
  Lin, and D.~McGlinchey, ``Minimal conditions for collectivity in
  ${e}^{+}{e}^{\ensuremath{-}}$ and $p+p$ collisions'',
  \href{http://dx.doi.org/10.1103/PhysRevC.97.024909}{{\em Phys. Rev. C}
  {\bfseries 97} (2018) 024909},
  \href{http://arxiv.org/abs/1707.02307}{{\ttfamily arXiv:1707.02307
  [nucl-th]}}.

\bibitem{Castorina:2020iia}
P.~Castorina, D.~Lanteri, and H.~Satz, ``{Strangeness enhancement and flow-like
  effects in $e^+e^-$ annihilation at high parton density}'',
  \href{http://dx.doi.org/10.1140/epja/s10050-021-00393-z}{{\em Eur. Phys. J.
  A} {\bfseries 57} (2021) 111},
  \href{http://arxiv.org/abs/2011.06966}{{\ttfamily arXiv:2011.06966
  [hep-ph]}}.

\bibitem{CMS:2015fgy}
{\bfseries CMS} Collaboration, V.~Khachatryan {\em et~al.}, ``{Measurement of
  long-range near-side two-particle angular correlations in pp collisions at
  $\sqrt s =$13 TeV}'',
  \href{http://dx.doi.org/10.1103/PhysRevLett.116.172302}{{\em Phys. Rev.
  Lett.} {\bfseries 116} (2016) 172302},
  \href{http://arxiv.org/abs/1510.03068}{{\ttfamily arXiv:1510.03068
  [nucl-ex]}}.

\bibitem{OrtizVelasquez:2013ofg}
A.~Ortiz~Velasquez, P.~Christiansen, E.~Cuautle~Flores, I.~Maldonado~Cervantes,
  and G.~Pai\'c, ``{Color Reconnection and Flowlike Patterns in $pp$
  Collisions}'', \href{http://dx.doi.org/10.1103/PhysRevLett.111.042001}{{\em
  Phys. Rev. Lett.} {\bfseries 111} (2013) 042001},
  \href{http://arxiv.org/abs/1303.6326}{{\ttfamily arXiv:1303.6326 [hep-ph]}}.

\bibitem{Bierlich:2022pfr}
C.~Bierlich {\em et~al.}, ``{A comprehensive guide to the physics and usage of
  PYTHIA 8.3}'', \href{http://dx.doi.org/10.21468/SciPostPhysCodeb.8}{{\em
  SciPost Phys. Codeb.} {\bfseries 2022} (2022) 8},
  \href{http://arxiv.org/abs/2203.11601}{{\ttfamily arXiv:2203.11601
  [hep-ph]}}.

\bibitem{Song:2010mg}
H.~Song, S.~A. Bass, U.~Heinz, T.~Hirano, and C.~Shen, ``{200 A GeV Au+Au
  collisions serve a nearly perfect quark-gluon liquid}'',
  \href{http://dx.doi.org/10.1103/PhysRevLett.106.192301}{{\em Phys. Rev.
  Lett.} {\bfseries 106} (2011) 192301},
  \href{http://arxiv.org/abs/1011.2783}{{\ttfamily arXiv:1011.2783 [nucl-th]}}.
  [Erratum: Phys.Rev.Lett. 109, 139904 (2012)].

\bibitem{Pierog:2013ria}
T.~Pierog, I.~Karpenko, J.~Katzy, E.~Yatsenko, and K.~Werner, ``{EPOS LHC: Test
  of collective hadronization with data measured at the CERN Large Hadron
  Collider}'', \href{http://dx.doi.org/10.1103/PhysRevC.92.034906}{{\em Phys.
  Rev. C} {\bfseries 92} (2015) 034906},
  \href{http://arxiv.org/abs/1306.0121}{{\ttfamily arXiv:1306.0121 [hep-ph]}}.

\bibitem{ALICE:2022wpn}
{\bfseries ALICE} Collaboration, ``{The ALICE experiment -- A journey through
  QCD}'', \href{http://arxiv.org/abs/2211.04384}{{\ttfamily arXiv:2211.04384
  [nucl-ex]}}.

\bibitem{Aamodt:2008zz}
{\bfseries ALICE} Collaboration, K.~Aamodt {\em et~al.}, ``{The ALICE
  experiment at the CERN LHC}'',
  \href{http://dx.doi.org/10.1088/1748-0221/3/08/S08002}{{\em JINST} {\bfseries
  3} (2008) S08002}.

\bibitem{ALICE:2014sbx}
{\bfseries ALICE} Collaboration, B.~B. Abelev {\em et~al.}, ``{Performance of
  the ALICE Experiment at the CERN LHC}'',
  \href{http://dx.doi.org/10.1142/S0217751X14300440}{{\em Int. J. Mod. Phys. A}
  {\bfseries 29} (2014) 1430044},
  \href{http://arxiv.org/abs/1402.4476}{{\ttfamily arXiv:1402.4476 [nucl-ex]}}.

\bibitem{Skands:2014pea}
P.~Skands, S.~Carrazza, and J.~Rojo, ``{Tuning PYTHIA 8.1: the Monash 2013
  Tune}'', \href{http://dx.doi.org/10.1140/epjc/s10052-014-3024-y}{{\em Eur.
  Phys. J. C} {\bfseries 74} (2014) 3024},
  \href{http://arxiv.org/abs/1404.5630}{{\ttfamily arXiv:1404.5630 [hep-ph]}}.

\bibitem{Brun:1994aa}
R.~Brun, F.~Bruyant, F.~Carminati, S.~Giani, M.~Maire, A.~McPherson,
  G.~Patrick, and L.~Urban, ``{GEANT Detector Description and Simulation
  Tool}'',.
\url{https://cds.cern.ch/record/1082634}.

\bibitem{ALICE:2023lyr}
{\bfseries ALICE} Collaboration, S.~Acharya {\em et~al.}, ``{Multiplicity and
  event-scale dependent flow and jet fragmentation in pp collisions at
  $\sqrt{s}$ = 13 TeV and in p$-$Pb collisions at $\sqrt{s_\mathrm{NN}}$ = 5.02
  TeV}'', \href{http://dx.doi.org/10.1007/JHEP03(2024)092}{{\em JHEP}
  {\bfseries 03} (2024) 092}, \href{http://arxiv.org/abs/2308.16591}{{\ttfamily
  arXiv:2308.16591 [nucl-ex]}}.

\bibitem{ALICE:2021nir}
{\bfseries ALICE} Collaboration, S.~Acharya {\em et~al.}, ``{Long- and
  short-range correlations and their event-scale dependence in
  high-multiplicity pp collisions at $\boldsymbol{\sqrt{{\textit s}}}=13$
  TeV}'', \href{http://dx.doi.org/10.1007/JHEP05(2021)290}{{\em JHEP}
  {\bfseries 05} (2021) 290}, \href{http://arxiv.org/abs/2101.03110}{{\ttfamily
  arXiv:2101.03110 [nucl-ex]}}.

\bibitem{Bierlich:2017vhg}
C.~Bierlich, G.~Gustafson, and L.~L{\"o}nnblad, ``{Collectivity without plasma
  in hadronic collisions}'',
  \href{http://dx.doi.org/10.1016/j.physletb.2018.01.069}{{\em Phys. Lett. B}
  {\bfseries 779} (2018) 58--63},
  \href{http://arxiv.org/abs/1710.09725}{{\ttfamily arXiv:1710.09725
  [hep-ph]}}.

\end{thebibliography}\endgroup
